\documentclass{PoS}
\newcommand{\KK}{{$KK$}}
\newcommand{\KKMC}{\KK MC}

\title{HERWIRI2: Exponentiated Electroweak Corrections in a Hadronic Event Generator}

\ShortTitle{HERWIRI2}

\author{\speaker{Scott A. Yost}%
	\thanks{This work and its presentation were supported in part by 
D.o.E. grant DE-PS02-09ER09-01 and grants from The Citadel Foundation.}\\
        The Citadel\\
        Charleston, SC 29409, USA\\
        E-mail: \email{scott.yost@citadel.edu}}
\author{Valerie Halyo
        \thanks{Work supported in part by D.o.E grant DE-FG02-91ER40671.}\\
        Princeton University\\
        Princeton, NJ 08544, USA\\
        E-mail: \email{valerieh@princeton.edu}}
\author{Miroslav Hejna\\
        Princeton University\\
        Princeton, NJ 08544, USA\\
        E-mail: \email{mhejna@princeton.edu}}
\author{B.F.L.\ Ward%
        \thanks{Work supported in part by D.o.E. grant DE-FG02-09ER41600.}\\
        Baylor University
        Waco, TX 76798, USA\\
        E-mail: \email{bfl\_ward@baylor.edu}}

\abstract{Reaching the desired precision level for $W$ and $Z$ processes at 
the LHC will require a mixture of higher-order QCD and electroweak corrections. 
HERWIRI2 is a step in implementing QED $\otimes$ QCD exponentiation in a 
hadronic event generator. This program implements leading electroweak 
corrections and coherent exclusive exponentiation in a HERWIG environment. We 
discuss the status of the program, recent tests, and future developments.
}

\FullConference{36th International Conference on High Energy Physics,\\
		July 4-11, 2012\\
		Melbourne, Australia}

\begin{document}

Precise measurements of vector boson production at the LHC will be very 
important in rigorous testing of the Standard Model. In addition, these 
production processes have been identified as standard candles for
the measurement of the beam luminosity.\cite{candle} The desired precision 
on the theoretical contribution to the error of these measurements is at the 
1\% level.  Recent studies \cite{EWsys} by some of the authors have
found that the state-of-the-art precision tag on single $Z$ production at 
a CMS energy of 10 TeV is $\sim 4.6\%$, of which approximately 2\% is due to
electoweak corrections. 
These studies were based on FEWZ\cite{fewz}, which provides NNLO QCD 
corrections, and HORACE, \cite{horace} which provides ${\cal O}(\alpha)$ 
radiative corrections with a final-state photon shower, and
PHOTOS, \cite{photos} which adds final state photonic radiation.

Attaining 1\% precision will require ${\cal O}(\alpha_s^2)$ (NNLO) QCD 
corrections, together with ${\cal O}(\alpha_s\alpha)$ electroweak corrections 
to next-to-leading log, and ${\cal O}(\alpha^2)$ to leading log. A general
framework based on generalized Yennie-Frautschi-Suura (YFS) 
exponentiation \cite{YFS} has been proposed to implement these corrections
incrementally in a hadronic event generator which should inherit some of the
advantages of YFS-exponentiated Monte Carlo programs developed for LEP 
physics, such as BHLUMI \cite{bhlumi}, \KKMC\ \cite{KKMC}, and related 
programs\cite{YFSrelated}. This framework has been named 
HERWIRI\cite{QCDxQED,HERWIRI1}, for ``High Energy Radiation With Infra-Red 
Improvements,'' and generalizes the YFS approach to encompass both 
QED and QCD exponentiation simultaneously.

The name HERWIRI acknowledges that the initial versions build
upon the HERWIG \cite{HERWIG} parton shower generator.  
The first to be released, HERWIRI1, \cite{HERWIRI1} implemented IR-improved 
splitting kernels \cite{kernels} obtained using the QCD analog of YFS 
exponentiation.  The IR-improved kernels have also 
been implemented \cite{HERWIRI1NLO} in MC$@$NLO \cite{MCNLO}.
The ultimate goal is a complete shower generator based entirely on 
QCD$\otimes$QED exponentiation with exact ${\cal O}(\alpha_s^2, \alpha_s 
\alpha, \alpha^2)$ residuals. \cite{QTS3-RADCOR05}

For electron-positron colliders, precision electroweak corrections have 
been implemented in the program \KKMC\ \cite{KKMC}, which had a 
precision tag for LEP2 of 0.2\%.
\KKMC\ uses YFS \cite{YFS} exponentiated multiple-photon 
radiation for both the initial and final state, and includes  
${\cal O}(\alpha)$ electroweak corrections \cite{ZF1, ZF2, ZF3}
via the DIZET6.21 \cite{DIZET} package developed for ZFITTER\cite{ZFITTER}. 
YFS residuals are calculated perturbatively to the relevant orders in 
$\alpha^k L^l$  ($L = \ln(s/m_e^2)$) and exact collinear 
bremsstrahlung is implemented for up to three photons. 

HERWIRI2\cite{radcorHW2} 
implements the electroweak radiative corrections of \KKMC\ in a 
hadronic shower generator, presently taken to be HERWIG. 
\KKMC\ benefits from a very efficient
representation of $n$-photon phase space, with complete control over the 
soft and collinear singularities for an arbitrary number of photons 
Real and virtual IR singularities cancel exactly to all orders. 
 
The Drell-Yan cross section with multiple-photon emission
can be expressed as an integral
over the parton-level process $q_i(p_1) {\overline q}_i(p_2)\rightarrow 
f(p_3) {\overline f}(p_4) + n\gamma(k)$, integrated over phase space and
summed over photons.  The parton momenta $p_1, p_2$ are generated using
parton distribution functions giving a process at CMS
energy $q$ and momentum fractions $x_1, x_2$ such that $q^2 = x_1 x_2 s$:
\begin{equation}
\sigma_{\rm DY} = \int {dx_1\over x_1} {dx_2\over x_2}  
        \sum_i 
        f_i(q, x_1)f_{\overline i}(q, x_2)
         \sigma_i (q^2)\delta(q^2-x_1 x_2 s),
\end{equation}
where the final state phase space includes $p_3, p_4$ and $k_i$, $i=1,\cdots, 
n$ and multiple gluon radiation + hadronization is included through a shower.

HERWIRI2 uses HERWIG 6.5\cite{HERWIG} as the shower generator, 
which creates the hard process first at Born level.
HERWIRI2 finds the $Z/\gamma^*$ and the partons interacting with it in the
event record.  The initial partons define $p_1$, $p_2$, which are transformed
to the CM frame and projected on-shell to create a starting point for \KKMC,
which generates the final fermion momenta $p_3, p_4$ and photons $k_i$ (both
ISR and FSR.) The generated particles are transformed back to the lab frame
and placed in the event record.
 
With a change of variables, the Drell-Yan cross section in HERWIG, and 
thus in HERWIRI2, can be expressed as
\begin{eqnarray}
\sigma_{\rm DY} &=& \int {dx_1\over x_1} {dx_2\over x_2}  \sum_i 
        f_i(q, x_1)f_{\overline i}(q, x_2)
         \sigma_i (q^2)\delta(q^2-x_1 x_2 s)\nonumber\\
&=& \int_{q_{\rm min}}^{q_{\rm max}} dq P(q) \int_{q^2/s}^1 {dx_1\over x_1} 
      \sum_i P_i\ W_{\rm HW}^{(i)}(q^2,x_1)
= \left\langle W_{\rm HW}\right\rangle
\end{eqnarray}
where $P(q)$ is a normalized, integrable, crude probability distribution for
$q$, $P_i$ is the crude probability of generating parton $i$, and
$W_{\rm HW}$ is the HERWIG event weight.  This weight depends only
on the hard Born cross section and is not altered by the shower.

The crude probability distributions used by HERWIG are
\begin{equation}
P(q) = \frac{1}{2}[P_\gamma(q) + P_Z(q)], \qquad 
P_\gamma(q)= {N_\gamma\over q^4}, \qquad
P_Z(q) = {N_2 q\over (q^2 - M_Z^2) + \Gamma_Z^2 M_Z^2}
\end{equation}
The HERWIG event weight is
\begin{equation}
W_{\rm HW} = \sum_i W_{\rm HW}^{(i)}, \qquad W_{\rm HW}^{(i)} = {1\over P(q)} 
        f_i(q, x_1)f_{\overline i}(q, x_2) \ln\left(s\over q^2\right) 
\sigma_{\rm HW}^{(i)}(q^2)
\end{equation}
and the corresponding probability for selecting parton $i$ is
\begin{equation}
P_i = W_{\rm HW}^{(i)}/W_{\rm HW}
\end{equation}

Electroweak corrections may be introduced via a form factor
\begin{equation}
F^{(i)}_{EW}(q^2) = {\sigma_i(q^2)\over \sigma_{\rm Born}^{(i)}(q^2)}
\label{EWffac}
\end{equation}
\KKMC\ will calculate the EW form factor, and multiply it by the 
HERWIG Born cross section. To avoid double-counting EW effects, any EW
parameters in the denominator of eq.\ (\ref{EWffac}) must match those in
HERWIG. The total cross section may be expressed as the average of a combined
weight, 
\begin{equation}
\sigma_{\rm tot} = \left\langle W_{\rm tot}\right\rangle , \qquad
W_{\rm tot} = F_{EW}^{(i)}(q^2) W_{\rm HW} = W_{\rm HW} 
{\sigma_{\rm KK}^{(i)}(q^2) \over \sigma_{\rm Born}^{(i)}(q^2)}.
\end{equation}

The \KKMC\ cross section is calculated using a primary distribution
\begin{equation}
{d\sigma_{\rm Pri}^{(i)}(s,v) \over dv} = 
 \sigma_{\rm Born}^{(i)}(s(1-v)) 
{1\over 2}\left(1 + {1\over{\sqrt{1-v}}}\right)
{\overline\gamma}_i v^{{\overline\gamma}_i-1} v_{\rm min}^{\gamma_i -
                        {\overline\gamma}_i}
\label{priDist}
\end{equation}
with
\begin{equation}
\gamma_i = {2\alpha\over\pi}Q_i^2 
\left[\ln \left({s\over m_i^2}\right)  - 1\right],
\qquad {\overline\gamma}_i 
        = {2\alpha\over\pi}Q_i^2\ln\left({s\over m_i^2}\right)
\end{equation}
to generate the factor $v$ giving the fraction of $s$ remaining after ISR 
photon emission, $s_X = s(1 - v)$.

The \KKMC\ cross section is
\begin{equation}
\sigma(q^2) = 
\int d\sigma_{\rm Pri} {d\sigma_{\rm Cru} \over d\sigma_{\rm Pri}}
{d\sigma_{\rm Mod}\over d\sigma_{\rm Cru}} 
= \sigma_{\rm Pri} \left\langle W_{\rm Cru} W_{\rm Mod}
\right\rangle .
\end{equation}
$W_{\rm Cru}$ is calculated during ISR generation and
$W_{\rm Mod}$ is generated after $s_X$ is available.

The HERWIG and \KKMC\ weights are combined to calculate the total 
HERWIRI2 weight, 
\begin{equation}
\sigma_{\rm tot} = \left\langle W_{\rm HW} 
        {\sigma_i(q^2)\over \sigma_{\rm Born}^{(i) \star}(q^2)}\right\rangle 
= \left\langle W_{\rm HW} \sigma_{\rm Pri}^{(i)}(q^2) 
        {W_{\rm Cru}^{(i)} W_{\rm Mod}^{(i)} 
        \over \sigma_{\rm Born}^{(i) \star}(q^2) }
\right \rangle, 
\end{equation}

This average will eventually be calculated using
a joint probability distribution for $q$ and $v$,
$D(q,v) = P(q) d\sigma_{\rm Pri}/dv$, with $P(q)$ from HERWIG.
An adaptive MC (S. Jadach's FOAM \cite{FOAM}) will calculate
the normalization of the distribution at the beginning of the run, in a
similar manner to how \KKMC\ presently integrates the one-dimensional
primary distribution.  However, as a first step, we have constructed
a version of HERWIRI2 using \KKMC's one-dimensional primary distribution. 

In the present scheme, the built-in primary distribution for electrons at 
scale $q_0 = M_Z$ is be used for the low-level generation of $v$.
The transformation from this distribution to a distribution at HERWIG's
generated scale $q$ for quark $i$ is then obtained by a change of variables.
The result may be expressed as an average of a product 
\begin{equation}
\sigma_{\rm tot} = \left\langle W_{\rm HW} W_{\rm Mod} W_{\rm Karl} W_{\rm FF}
W_{\gamma} \right\rangle
\end{equation}
with new weights defined by
\begin{equation}
W_{\rm Karl} = \frac{\sigma_{\rm Pri}^{(e)} W_{\rm Crud}^{(i)}}{
                \sigma_{\rm Born}^{(e)}(q_0^2(1-v))} ,\qquad
W_{\rm FF} = \frac{\sigma_{\rm Born}^{(i)}(q^2(1-v))}{ 
                \sigma_{\rm Born}^{(i) \star} (q^2)}, \qquad
W_\gamma = \frac{{\overline\gamma}_i}{{\overline\gamma}_e}
        \frac{F_{\rm YFS}^{(i)}}{F_{\rm YFS}^{(e)}} v^{\gamma_i - \gamma_e},
\end{equation}
with YFS form factors
\begin{equation}
F_{\rm YFS}^{(i)} = \frac{e^{-C_E\gamma_i}}{\Gamma(1+\gamma_i)},\qquad
F_{\rm YFS}^{(e)} = \frac{e^{-C_E\gamma_e}}{\Gamma(1+\gamma_e)},
\end{equation}
and Euler's number $C_E=0.5772...$.
The $\gamma$ factors are calculated using $q^2/m_i^2$ for parton $i$ and
$q_0^2/m_e^2$ for the electron. The weight $W_\gamma$ has been modified
since the first publication on HEWIRI2\cite{radcorHW2}, and may be modified
further, due to the discovery of some uncancelled dependence on $v_{\min}$, 
a cutoff in eq.\ (\ref{priDist}) which should not affect the final result.

HERWIRI2 is still under development, so any numerical results must be treated 
as preliminary.  
A $10^6$-event run for $pp$ collisions at 8 TeV with the
$Z/\gamma^*$ invariant mass bounded by 30 GeV and 300 GeV, using HERWIG 6.520 
default parameters and CT10 PDFs \cite{CT10}, yields
a cross-section of $1218\pm 13$ pb, which is a 5.9\% electroweak correction,
a reasonable magnitude in light of previous calculations.\cite{EWsys}
An average of 0.45 ISR photons and 0.61 FSR photons are generated per event, with average total energies of 0.63 and 1.16 GeV, respectively.
 
Work is in progress to optimize MC generation in the presence of ISR.
As noted above, there is still some residual dependence on a cutoff 
$v_{\min}$ in \KKMC, which will require further refinement of the weights.
This can be traced to the fixed scale $q_0$ in generating 
the primary distribution for ISR.  The best solution of this will probably 
be to use \KKMC's beamsstrahlung 
feature to better model the range of parton CMS energies generated by HERWIG.
It will be especially interesting to see the effect of initial state 
radiation, which appears to enter at the 2 -- 3\% level, making it crucial to 
precision calculations.  HERWIRI2 will be an important step
toward the goal of a hadronic event generator based on nonabelian 
QCD$\otimes$QED exponentiation with exact  ${\cal O}(\alpha_s^2, 
\alpha_s \alpha, \alpha^2)$ residuals.

\section*{Acknowledgments}
S. Yost thanks the organizers of ICHEP 2012 for the invitation to present these
results, and D. Marlow and the Princeton Physics Department for their 
support and hospitality during a critical period of its development. S. Yost 
and B.F.L. Ward also acknowledge the hospitality of the CERN theory division,
which contributed greatly to the progress of this work.

\end{document}